\documentclass[10pt,a4paper]{article}
\usepackage[latin1]{inputenc}
\usepackage{amsmath}
\usepackage{amsfonts}
\usepackage{amssymb}
\usepackage{enumerate}
\usepackage{verbatim}
\usepackage{multirow}
\usepackage{amssymb}
\usepackage{stmaryrd}
\usepackage{array}
\usepackage[T1]{fontenc}
\usepackage[normalem]{ulem}
\usepackage[english]{babel}    
\usepackage[english]{layout} 
\usepackage{verbatim}
\usepackage[pdftex]{graphicx}
\usepackage{pgf}
\usepackage{color}

\newcommand{\degree}{$^{\rm o}$ }

\newcommand{\s}{$\sqrt{s} = 7\ $}
\newcommand{\pt}{\ensuremath{p_{\rm t}} }
\newcommand{\dndpt}{d$N$/d\ensuremath{p_{\rm t}} }

\begin{document}

\begin{center}


{\Large\bf Heavy flavour decay muon production at forward rapidity in pp collisions at \s TeV and in Pb\--Pb collisions at $\sqrt{s_{\mathrm{NN}}} = 2.76$\ TeV with the ALICE experiment}\\

\vspace*{0.4cm}

{\Large L.~Manceau for the ALICE Collaboration}\\
INFN and Universit\'{a} di Torino, Via P. Guiria 1, I-10125 Torino (Italy)\\
\vspace*{0.1cm}
E-mail: lmanceau@to.infn.it
\end{center}
\vspace*{0.15cm}

\noindent {\bf Abstract.} The production of muons from heavy flavour decays is measured at forward rapidity ($2.5<y<4$) with the muon spectrometer of the ALICE experiment at the LHC. In pp collisions at \s TeV, the measurement of the transverse momentum and rapidity differential production cross section of muons from heavy flavour decays is carried out from a data sample corresponding to an integrated luminosity of $L_{\mathrm{int}}=16.5\ \mathrm{nb^{-1}}$ and results are compared to predictions based on perturbative QCD calculations. In Pb\--Pb collisions at $\sqrt{s_{\mathrm{NN}}} = 2.76$\ TeV, the analysis of a data sample corresponding to an integrated luminosity of $L_{\mathrm{int}}=2.48\ \mathrm{\mu b^{-1}}$ provides a measurement of the ratio of inclusive muon yield in central to peripheral collisions ($R_{\mathrm{CP}}$) as a function of the collision centrality.


\section{Introduction}

The unprecedented regime of energy reached at the LHC has opened a new era for the study of deconfined QCD matter expected to be created in the extreme thermodynamical conditions characterising early stages of heavy ion collisions. Heavy flavours (charm and beauty) are abundantly produced at LHC energies and their measurement should bring essential information on deconfined QCD matter as they are expected to be produced during the initial hard scattering and to coexist with the surrounding medium due to their long life time~\cite{Alessandro:2006yt}. In particular, heavy quarks are expected to lose energy while passing through deconfined QCD matter~\cite{Dokshitzer:2001zm} and open heavy flavour production measurement should help to probe the system density. 
It is important to note that the study of heavy flavours in heavy ion collisions requires a baseline provided by their measurement in pp collisions.

ALICE (A Large Ion Collider Experiment)~\cite{KAaJINST} is the LHC experiment dedicated to the study of heavy ion collisions. At forward rapidity ($2.5<y<4$), ALICE is equipped with a spectrometer~\cite{KAaJINST} allowing for the measurement of heavy flavours via their muonic decay channels. It is composed of a passive frontal absorber made of a combination of low and high atomic number material, a beam shield, a $3$ T m dipole magnet, five tracking stations and two trigger stations behind a $1.2$ m thick iron wall. 

In 2010 ALICE has collected data in pp (Pb\--Pb) collisions at \s TeV ($\sqrt{s_{\mathrm{NN}}} = 2.76$\ TeV). Results on the measurement of single muons from heavy flavours in pp collisions are presented and compared to predictions based on perturbative QCD. In addition, results on the ratio of inclusive muon yield in central to peripheral Pb\--Pb collisions as a function of the collision centrality are reported.   

\section{Results in pp collisions at \s TeV}
\label{pp}

The single muon analysis in pp collisions at \s TeV is described in detail in~\cite{Abelev:2012pi}. It uses a data sample corresponding to an integrated luminosity of  $L_{\mathrm{int}}=16.5\ \mathrm{nb^{-1}}$. Events in the data sample contain at least one reconstructed muon with a transverse momentum (\ensuremath{p_{\rm t}}) above a threshold of $\sim 0.5$ GeV/c.  The beam induced background is reduced by means of an offline selection. The events selected for the analysis are required to be associated with an interaction vertex reconstructed with the ALICE central barrel detectors. Selected tracks in the muon spectrometer have a reconstructed angle at the end of the absorber between $171$\degree and $178$\degree and pass a sharp cut corresponding to the acceptance of the spectrometer ($-4<\eta<-2.5$). In addition, the track candidate in the tracker system has to match with a track reconstructed in the trigger system and has to point to the pp interaction vertex. The latter condition is particularly efficient at removing the remaining beam induced background and the remaining tracks with mis-associated hits.

After track selection, the background for the heavy flavour decay signal is mainly composed of muons from primary light hadrons (mainly pions and kaons) and of muons from secondary light hadrons produced by hadronic interactions in the frontal absorber. 
The analysis is focused on \pt larger than $2$ GeV/c. In this range, muons from secondary light hadron decay are estimated from simulation to be close to negligible ($\sim3\%$) while muons from primary hadrons represent the dominating source of background ($\sim 25\%$). Background subtraction was performed by means of simulations with the PYTHIA~\cite{Sjostrand:2004ef} and PHOJET~\cite{Engel:1995sb} event generators. The normalisation is done assuming that, for \pt$<1$ GeV/c, the fraction of background in the data is the same as in the simulation. Finally, the mean of the fitted \pt distributions from the PYTHIA and PHOJET event generators is subtracted from the measured \pt distribution.

After background subtraction, yields are corrected for acceptance and efficiency ($\sim 90\%$)  and converted into a muon cross section using the minimum bias pp cross section derived from a van der Meer scan ($\sigma_{MB}=62.5\pm2.2\ (syst.)$, statistical uncertainty is negligible)~\cite{MGKA}. The systematic uncertainty on muon cross section varies between $8-14\%$ in the rapidity range $2.5<y<4$ and is dominated by the systematics from background subtraction. 

 \begin{figure}[h]
\centering
\begin{center}
\includegraphics[width=12.cm]{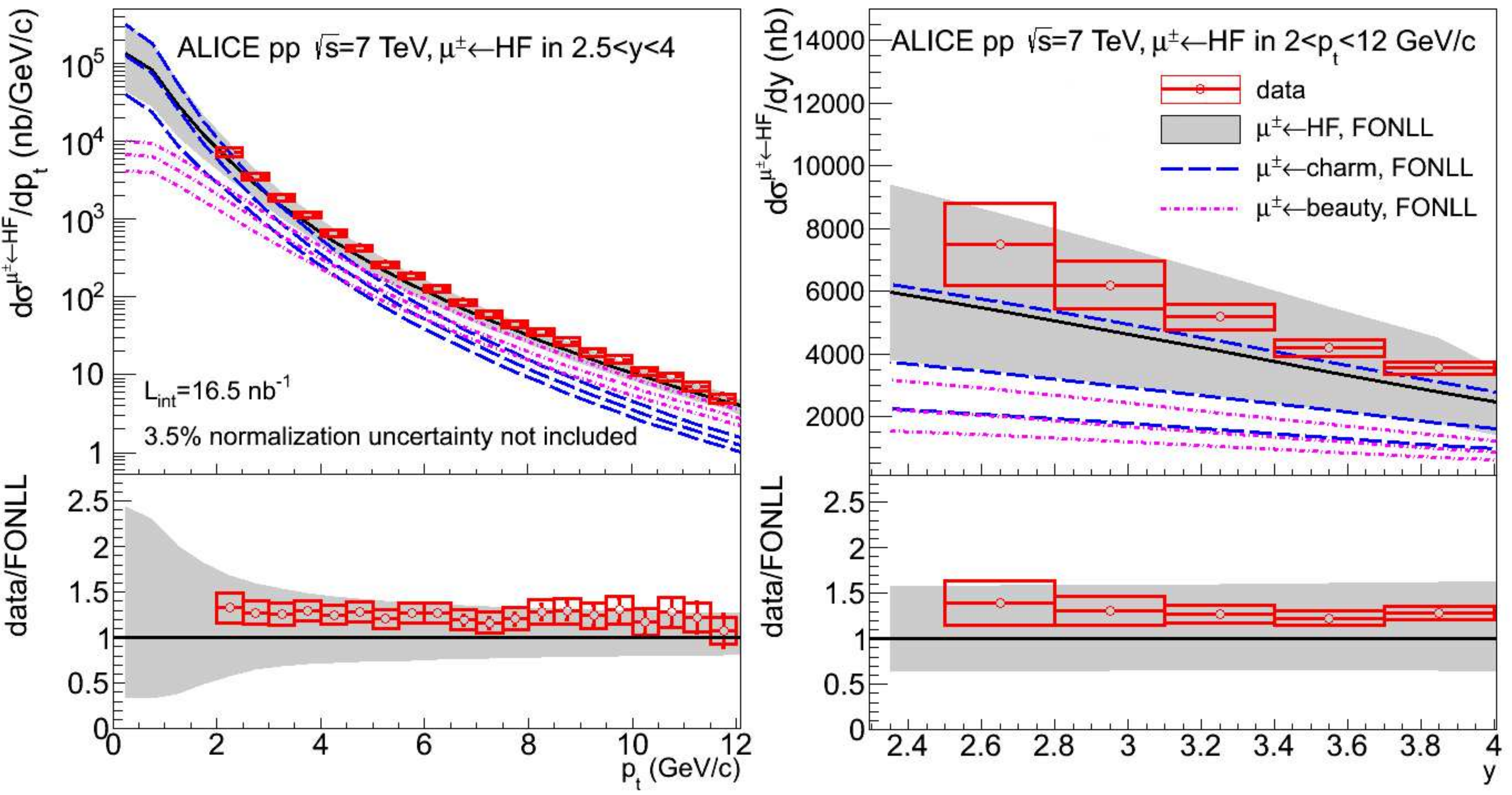}
\caption{Inclusive \ensuremath{p_{\rm t}}-differential (left) and $y$-differential (right) production cross section of muons from charm and beauty quark decay~\cite{Abelev:2012pi}.}
\label{pp7TeV}
\end{center}
\end{figure}

Figure~\ref{pp7TeV} shows the measured production differential cross section of muons from charm and beauty quark decay as a function of \pt (left) and $y$ (right). Error bars and boxes correspond to statistical and systematic uncertainties, respectively. The systematic uncertainty on $\sigma_{MB}$ is not displayed. Measurements are compared to Fixed-Order Next-to-Leading Log (FONLL) perturbative QCD predictions~\cite{Cacciari:2001td}. Bottom panels of Fig.~\ref{pp7TeV} display deviations between peturbative QCD predictions and measurements. A good agreement is observed within uncertainties. It worth noting that according to FONLL perturbative QCD predictions muons from beauty quark decays should dominate for \pt larger than $6$\ GeV/c.   

\section{Results in Pb\--Pb collisions at $\sqrt{s_{\mathrm{NN}}} = 2.76$ TeV}

The analysis is based on a minimum bias data sample corresponding to an integrated luminosity of $L_{\mathrm{int}}=2.48\ \mathrm{\mu b^{-1}}$. Events are selected according to their degree of centrality by means of the VZERO detector~\cite{KAaJINST} amplitude which is fitted using the Glauber model~\cite{Collaboration:2011rta} to determine intervals in percentages of the nuclear cross section. An offline selection aiming at reducing beam induced background and fake tracks is performed in a very similar way as for the pp collision analysis (see section~\ref{pp}). The background of muons from light hadron decay is not subtracted. The analysis is focused on \pt larger than $6$ GeV/c as, according to a HIJING~\cite{Wang:1991hta} generator simulation (without quenching), the background in this \pt range should not exceed $2-9\%$ depending on the centrality of the collisions. The data are corrected for acceptance and efficiency and the nuclear modification factor of inclusive muons in central to peripheral collisions ($R_{\mathrm{CP}}$) is determined as:

\begin{center}
\begin{equation}
R_{\mathrm{CP}}=\frac{\left[1/T_{\mathrm{AA}}\times \mathrm{d}N/\mathrm{d}p_{\mathrm{t}}\right]_{central}}{\left[1/T_{\mathrm{AA}}\times \mathrm{d}N/\mathrm{d}p_{\mathrm{t}}\right]_{peripheral\ (60-80\%)}},
\label{RCP}
\end{equation}
\end{center}

\noindent where $T_{\mathrm{AA}}$ is the nuclear overlap function from Glauber model~\cite{Collaboration:2011rta} and \dndpt is the muon \pt distribution.

 \begin{figure}[h]
\centering
\begin{center}
\includegraphics[width=8.5cm]{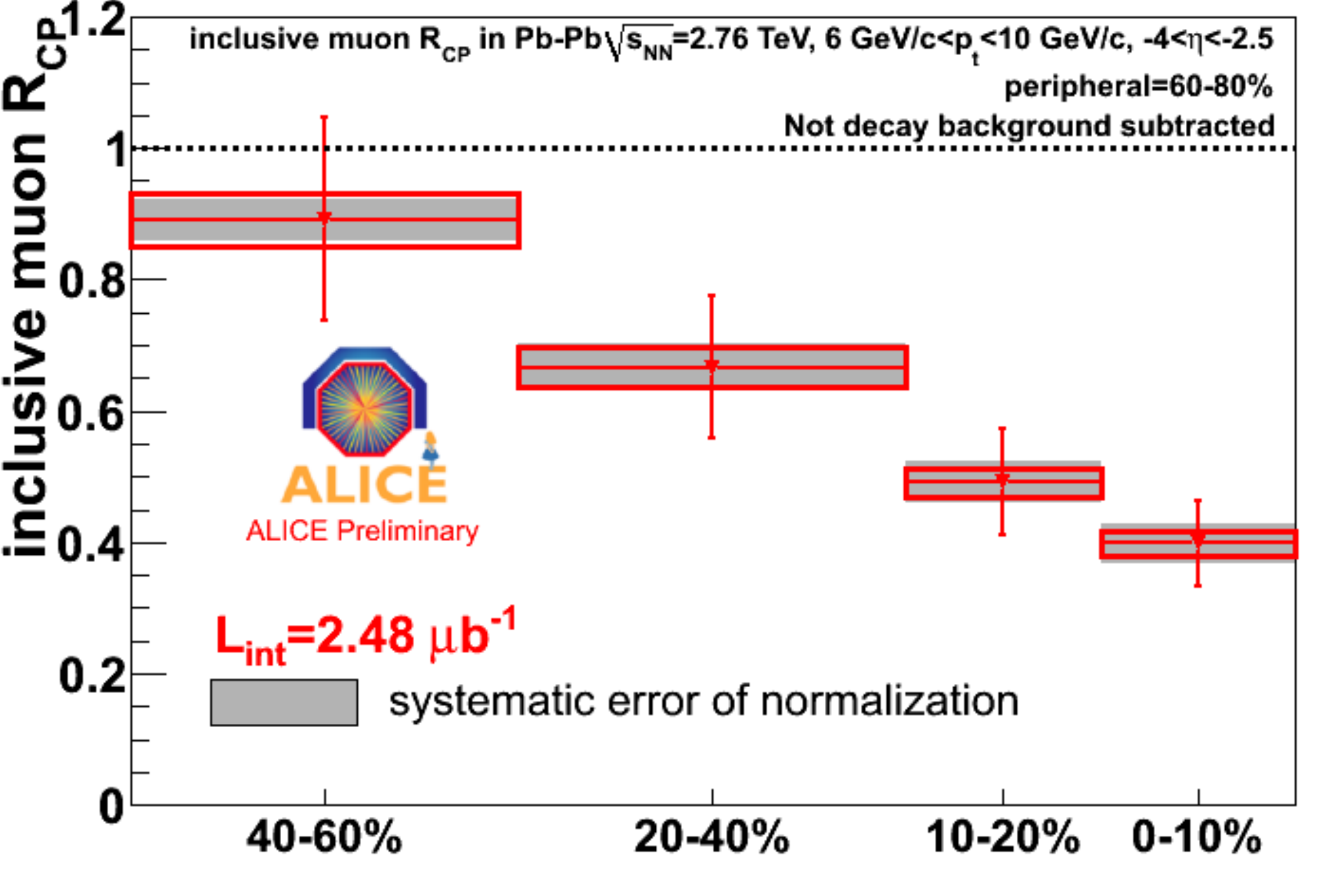}
\caption{Inclusive muon $R_{\mathrm{CP}}$ as a function of the centrality in the \pt range $6-10$ GeV/c.}
\label{PbPb276TeV}
\end{center}
\end{figure}

Figure~\ref{PbPb276TeV} presents the inclusive muon $R_{\mathrm{CP}}$ as a function of the collision centrality in the \pt range $6-10$ GeV/c. A suppression of muon production is observed and increases with centrality. This trend indicates that heavy quarks have lost energy in the medium created in the collisions. This observation is compatible with the formation of a strongly interacting medium.

\section{Conclusion}

The differential production cross section of muons from heavy flavour decay as a function of \pt and $y$ was measured in pp collisions at \s TeV in the acceptance range $2.5<y<4$ and in the \pt range $2<$ \pt$<10$ GeV/c. The FONLL perturbative QCD predictions provide a good description of the measurement.
In Pb\--Pb collisions at $\sqrt{s_{\mathrm{NN}}} = 2.76$ TeV, a suppression of muons from heavy flavour decays increasing with the centrality of the collision was observed. This trend indicates the formation of a strongly interacting medium.



\begin{thebibliography}{} 
\bibitem{Alessandro:2006yt}
  B. Alessandro, (Ed.) {\it et al.} [ ALICE Collaboration ],
  ``ALICE: Physics performance report, volume II,''
  J.\ Phys.\ G {\bf G32 } (2006)  1295-2040.
\bibitem{Dokshitzer:2001zm}
  Y.~L.~Dokshitzer and D.~E.~Kharzeev,
  Phys.\ Lett.\ B {\bf 519} (2001) 199,
  [hep-ph/0106202].
\bibitem{KAaJINST}
K.~Aamodt {\it et al.}  [ALICE Collaboration],
JINST\ 3,\ S08002\ (2008).
\bibitem{Abelev:2012pi}
  B.~Abelev {\it et al.}  [ALICE Collaboration],
  Phys.\ Lett.\  B {\bf 708} (2012) 265,
  [arXiv:1201.3791 [hep-ex]].
\bibitem{Sjostrand:2004ef}
  T.~Sjostrand, P.~Z.~Skands,
  Eur.\ Phys.\ J.\  {\bf C39 } (2005)  129-154,
  [hep-ph/0408302].
\bibitem{Engel:1995sb}
  R.~Engel, J.~Ranft and S.~Roesler,
  Phys.\ Rev.\  D {\bf 52} (1995) 1459,
  [arXiv:hep-ph/9502319].
\bibitem{MGKA}
M.~Gagliardi {\it et al.} [ALICE Collaboration], [arXiv:1109.5369 [hep-ex]].
\bibitem{Cacciari:2001td}
  M.~Cacciari, S.~Frixione and P.~Nason,
  JHEP {\bf 0103} (2001) 006,
  [arXiv:hep-ph/0102134].
\bibitem{Collaboration:2011rta}
  A.~Toia,
  J.\ Phys.\ G {\bf 38} (2011) 124007,
  [arXiv:1107.1973 [nucl-ex]].
\bibitem{Wang:1991hta}
  X.~N.~Wang and M.~Gyulassy,
  Phys.\ Rev.\  D {\bf 44} (1991) 3501.

\end{thebibliography}
\end{document}